\documentstyle[12pt]{article}

\title{\bf Gravitational complementary principle:\\ A new approach to quantum gravity }
\author{F. Darabi \thanks{e-mail:f.darabi@azaruniv.edu} \\
{\small Department of Physics, Azarbaijan University of Tarbiat
Moallem, 53714-161 Tabriz, Iran .} }
\begin{document}
\maketitle
\begin{abstract}
A new idea of quantum gravity is developed based on {\it
Gravitational Complementary Principle}. This principle states that
gravity has dual complement features: The quantum and classical
aspects of gravity are complement and absolutely separated by the
planck length into planckian and over-planckian domains,
respectively. The classical Einstein equations are correct at the
fundamental level at over-planckian domain and general relativity
is not a low energy limit of a more fundamental theory. The
quantum gravity is totally confined to the planckian domain with a
new kind of ultra-short range interaction, mediated by massive
(Planck mass) particles, through the virtual microscopic wormholes
of the Planck scale with action $\hbar$. There is no room for
gravitons or extra dimensions in this scenario. It is shown that
the hierarchy problem can solve the cosmological constant problem
via this new quantum gravity.
\end{abstract}

\newpage

\section{Introduction}

In the past few decades remarkable progress has been put forwarded
in constructing a unified theory of the forces of nature.
Weinberg-Salam theory of electroweak interaction has unified the
electromagnetic and weak interactions \cite{WS}, while the
so-called grand unified theories incorporates the strong
interaction into a wider gauge theory $SU(5)$. However, gravity as
the odd one out of four forces of nature resists against a
consistent formulation in the quantum framework. Combining general
relativity with quantum mechanics is the last hurdle to be
overcome in the "quantum revolution" \cite{Gi}. Many different and
powerful ( perturbative and non-perturbative ) approaches have
been pursued to quantize gravity, but as yet a completely
satisfactory quantum gravity theory remains elusive \cite{Ro}.
String and superstring theories, on the other hand, as the best
candidate to formulate a full quantum theory of gravity, have not
yet proceed successfully in this direction \cite{Ro}. The quantum
formulation of gravity involves some fundamental conceptual and
technical difficulties. The dual role played by the field $g_{\mu
\nu}$ as both the background spacetime structure and the quantity
which describes the dynamical aspect of quantum gravity causes the
entire notion of causality to be ill-defined. On the other hand,
the quantum theory of gravity is non-renormalizable due to the
dimensionality of the gravitational coupling constant.

From other points of view, combining general relativity with
quantum mechanics is inconsistent. General relativity demands
four-dimensional covariance while quantum mechanics is a theory of
measurement on 3-dimensional space-like hypersurfaces.

In spite of these technical difficulties and conceptual
complications it is generally believed that Einstein's general
relativity is a low energy limit of a more fundamental {\it
background independent} quantum theory of gravity \cite{Ro},
\cite{Wa}. It is discussed that the spacetime metric is coupled to
matter sources through Einstein equations and quantum theory
applies to these matter sources. Thus, any attempt to treat
gravity classically leads to serious problems and the only way to
avoid these difficulties is to treat the spacetime metric in a
probabilistic fashion which means quantum gravity.

In what follows, we will try to develop a new idea of quantum
gravity which may shed light on the current problems in quantizing
gravity, cosmological constant and hierarchy problems. We
emphasize that this idea is still far from being complete, but it
is hoped that it can stimulate further investigations in this
direction.

\section{Gravitational complementary principle}

The so-called complementary principle, introduced by Bohr, is one
of the most important basis of quantum mechanics. This fundamental
principle in mathematical language is equal to the Heisenberg
uncertainty principle which leads to the basic commutation
relations on which the standard quantum mechanics stands. In
quantum mechanics, position and momentum, just like particle
behavior and wave aspects of a system respectively, are complement
properties of the system and the theory {\it does not admit the
possibility of an experiment in which both could be established
simultaneously}. In this way, Bohr complementary principle BCP
resolved some serious difficulties in the advent of quantum
mechanics. The two slit experiment was one of confusing behaviors
of nature which was satisfactory described by this principle. The
prescription was to describe this experiment completely by the
wave aspect and avoid any particle behavior. In fact, any attempt
to describe it by the particle behavior would destroy the whole
interference pattern.

Inspired by Bohr complementary principle BCP and its key role in
the foundation of quantum mechanics we introduce {\it
Gravitational Complementary Principle }GCP:\\
Classical and quantum aspects of a gravitational system are
complement properties of the system and the gravity theory {\it
does not admit the possibility of a gravitational interaction in
which both classical and quantum aspects could be established
simultaneously}. A quantum mechanical experiment demands {\it
wave} or {\it particle} description according to the nature of
that experiment. For example, the two slit experiment needs wave
description whereas the photoelectric experiment demands particle
description. A gravitational interaction, however, demands {\it
classical} or {\it quantum} description according to the size (or
energy scale) of the domain in which that interaction takes
place\footnote{Similar ideas have already been proposed. For
example, in \cite{Pa}, the author speaks of a thermodynamic
spacetime and it is possible to separate the domain of validity of
classical and quantum gravitational effects while connecting them
through thermodynamic identities. }. We then introduce a
characteristic length scale for demarcation between the classical
and quantum features of gravitational interactions. Since GCP is
absolutely strict on the classical and quantum domains, then this
length scale must be a fundamental one to strictly forbid any
intrusion between these disjoint domains.

It was originally pointed out by Planck that combining the
universal constants $G, \hbar$ and $c$ gives a new fundamental
unit of length, namely the Planck length
$$
L_{pl}=\left(\frac{G\hbar}{c^3}\right)^{1/2}=1.616 \times 10^{-33}
cm.
$$
The Planck length should be thought of as the {\it zero point
length} of the spacetime and any correct theory of quantum gravity
must incorporate this feature in suitable form \cite{Pa}. This
fundamental length scale is generic to the gravitational
interactions, due to the presence of $G$, and there is no
analogous typical length scale associated with other three
interactions. This is a unique feature of gravitational
interaction so as to be distinguished from other interactions.
This tells us that gravity is different in nature from other
forces. The existence of this length scale in the framework of
gravitational interactions is completely consistent with GCP
provided we absolutely divide the classical and quantum features
of the gravitational interactions into disjoint {\it
over-planckian} and {\it planckian} domains, respectively. While
other three interactions obey the rules of BCP ( quantum mechanics
) at over-planckian domain, it seems gravity at this domain does
not. Rather, it obeys the rules of quantum mechanics at planckian
domain! At over-planckian domain it is just a pure classical field
theory.

GCP states that: unlike other three forces where a typical
interaction can be described both classical and quantum
mechanically by the same action principle, the gravitational
interaction with dual features at over-planckian and planckian
domains is described by two different kinds of interactions and
action principles in these domains. In other words, unlike for
example the electromagnetic interactions where $A_{\mu}$ plays
same ( dynamical ) role in classical and quantum descriptions, in
the case of gravitational interactions $g_{\mu \nu}$ does not play
the same role in classical and quantum descriptions. In fact,
$g_{\mu \nu}$ does not play a dynamical role in quantum
description. It merely can be described by the classical field
theory of gravity ( Einstein-Hilbert action ) at over-planckian
domain and has nothing to do with quantum features of gravity
which are trapped at planckian domain. Another interaction and
action principle is responsible for the quantum features of
gravitational interaction at planckian domain! The principles of
quantum theory do not apply to gravity at over-planckian domain
and the classical general relativity is correct at the fundamental
level. The classical Einstein's general relativity at
over-planckian domain is no longer a low energy limit of a more
fundamental theory of quantum gravity. One can not use $g_{\mu
\nu}$ as a quantum field to describe the quantum features of
gravity at over-planckian domain. The quantum features of gravity
are totally confined and trapped in the planckian domain and are
described by a field other than $g_{\mu \nu}$\footnote{This
property of quantum gravity is similar to the weak interactions in
which the massive gauge bosons are mediated in a small range.}.
Therefore, any attempt to study the quantum gravity is strictly
limited into the planckian domain and the resultant theory of
quantum gravity will be $g_{\mu \nu}$ independent. In this way, we
will hope to have a new quantum gravity without technical and
conceptual difficulties, discussed so far, arising due to dual
roles of $g_{\mu \nu}$ as classical background and quantum fields.
In fact, all above problems arise because we violate GCP by
assuming dual roles for $g_{\mu \nu}$, simultaneously.

Just like the two slit experiment where the assumption of particle
behavior (photon) for radiation violates BCP and destroys the
whole interference pattern, the assumption of quantum behavior
(graviton) for gravitation $g_{\mu \nu}$ at over-planckian domain
violates GCP and destroys the whole quantization pattern, for
example by the appearance of non-renormalizable divergences!

As a rule, a wrong question from nature yields a nonsense answer.
Most of the current approaches to quantum gravity obey this rule,
concerning GCP . In these approaches, usually a simultaneous
combination of classical and quantum features of gravity is used
to quantize gravity which violates GCP. For example, we know the
quantum version of Einstein equation
\begin{equation}
G_{\mu \nu}=8\pi G_{N}<T_{\mu \nu}>, \label{1}
\end{equation}
leads to a serious problem that the gravitational field behaves in
a discontinuous acausal way and this difficulty seems to be
avoided only by treating spacetime metric in a probabilistic
fashion \cite{Wa}. Using GCP, we show that this is not necessarily
the case. In fact, Eq.(\ref{1}) is generally a wrong question from
the system and so leads to the above problem. According to GCP,
the field $g_{\mu \nu}$ is a pure classical object at
over-planckian domain and has nothing to do with any effect in
which the observer may play the key role by the act of
observation. Therefore, $g_{\mu \nu}$ can in no way couple to a
matter source with some quantum probabilistic ( observer dependent
) features. One can resolve this confliction by assuming that any
source of matter which is quantized over 3-hypersurfaces, behaves
in a fully deterministic way against the 4-dimensional classical
object $g_{\mu \nu}$. In fact, for the given excited states
$|\alpha>$, $|\beta>$ the object $<\alpha|T_{\mu \nu}|\beta>$
implies the ignorance of 3-D observer about the full behavior of
the system under consideration, and impelling $g_{\mu \nu}$ as a
4-D object to obey this ignorance is not really justified. In
other words, every quantized non-gravitational system at
over-planckian domain which behaves in a probabilistic way for 3-D
observers, may behave against 4-D classical gravity in a fully
deterministic and observer independent way\footnote{This is in
spirit of Einstein's believe: {\it God does not play dice on
$g_{\mu \nu}$ background! }}. Eq.(\ref{1}) is a simultaneous
combination of classical ($g_{\mu \nu}$) and quantum probabilistic
($<\alpha|T_{\mu \nu}|\beta>$) notions in quantizing gravity at
over-planckian domain and so violates GCP. According to GCP, since
this is not a correctly written equation then it yields nonsense
answer, namely a discontinuous acausal behavior of the gravity. By
this ( wrong ) equation and its nonsense answer one may not
conclude that $g_{\mu \nu}$ should be quantized at over-planckian
domain. However, it is important to note that the vacuum
expectation value $<0|T_{\mu \nu}|0>$ is understood as a classical
( observer independent ) concept and so can couple to the geometry
through equation (\ref{1}). For example we know the mass of
fundamental particles are proportional to the Higgs field vacuum
expectation value $<0|\phi|0>$. On the other hand, these particles
can trivially couple to the geometry through classical Einstein
equation. Therefore, one concludes that $<0|\phi|0>$ must couple
to the geometry as well, through equation (\ref{1}).

In the covariant perturbation method also one writes the spacetime
metric $g_{\mu \nu}$ as
\begin{equation}
g_{\mu \nu}=\eta_{\mu \nu}+h_{\mu \nu}, \label{2}
\end{equation}
where $\eta_{\mu \nu}$ is a background spacetime and $h_{\mu \nu}$
represents a self-interacting spin-2 quantum field propagating on
this background. This approach leads to a non-renormalizable
perturbation theory \cite{Wa}. As is easily seen, in this approach
a simultaneous combination of classical and quantum field is
presented which apparently violates GCP. According to GCP, the
object $g_{\mu \nu}$ is intrinsically a classical field at
over-planckian domain and so can not be a simultaneous combination
of classical and quantum parts. Indeed, the non-renormalizability
of quantum gravity in this approach is a direct consequence of GCP
violation which leads the pattern for quantizing gravity to be
completely lost, in the same way as a wave interference pattern is
lost when one violates BCP by introducing the particle behavior in
the two slit experiment. Eq.(\ref{2}) is just correct at
over-planckian domain provided one does not interpret $h_{\mu
\nu}$ as a quantum field (graviton) in this domain. This equation
just describes a classical gravitational wave $h_{\mu \nu}$
propagating on the classical background $\eta_{\mu \nu}$. Since
gravitational interaction does not obey BCP ( wave-particle
duality ) at over-planckian domain, then there is no quantum
particle ( graviton ) associated with the gravitational wave,
$h_{\mu \nu}$.

In the canonical approach one attempts to construct a quantum
theory in which the Hilbert space carries a representation of the
operators corresponding to the full metric, or some functions of
the metric, without background metric to be fixed. This approach
also uses simultaneously the classical and quantum notions by
associating $g_{\mu \nu}$ with operator representation.

It seems in all other current approaches to quantum gravity one
can find simultaneous use of classical and quantum features of
gravity.

\section{Ultra-short range quantum gravity}

Up to know, we have introduced GCP which tells us that a
gravitational interaction at over-planckian domain is completely
described by classical general relativity with no notions of
quantum gravity, in principle. GCP learns us that at
over-planckian domain general relativity is not the low energy
limit of a fully quantized gravity. There is no quantum gravity
(graviton) at over-planckian $g_{\mu \nu}$ domain, at all. Quantum
gravity is absolutely confined and trapped at planckian domain
where $g_{\mu \nu}$ is not defined. If so, the important question
is: {\it what is the real nature and physics of quantum
gravitational interactions at planckian domain?}

To answer this question, one may look for quantum gravitational
interactions of ultra-short range at planckian domain. In so
doing, one may get suspicious to the following relations
\begin{equation}
M_{pl}^2=\hbar\frac{c}{G_{N}}, \label{4}
\end{equation}
\begin{equation}
L_{pl}^2=\hbar\frac{G_{N}}{c^3}. \label{4'}
\end{equation}
These relations are reminiscent of the well-known relations in
quantum mechanics
\begin{equation}
E=\hbar \omega, \label{5}
\end{equation}
\begin{equation}
P=\hbar K, \label{5'}
\end{equation}
which describe quantitatively the wave-particle duality or in a
sense BCP. Here, the Planck constant $\hbar$ relates the dual
features of nature, namely the wave aspects and particle behaviors
of a quantum mechanical system. Inspired by this key role of
$\hbar$ in BCP, one may think that it may play a similar role in
relating other dual features of nature, namely the classical
aspects and quantum behaviors of a gravitational system. In other
words, Eqs.(\ref{4}), (\ref{4'}) play the same role in GCP as
Eqs.(\ref{5}), (\ref{5'}) does in BCP. Therefore, since the
factors $\frac{c}{G_{N}}$ and $\frac{G_{N}}{c^3}$ in the r.h.s of
Eqs.(\ref{4}), (\ref{4'}) are pure classical quantities one may
think that $M_{pl}$ and $L_{pl}$ in the l.h.s must be quantities
representing the quantum features of the gravitational system.
Since $c$ and $G$ describe a classical {\it long range} ($c$),
{\it gravitational} ($G$) interaction, respectively, then $M_{pl}$
and $L_{pl}$ may be interpreted respectively as the particle's
mass and length scale of the quantum gravity
interaction\footnote{One may wish as well to interpret the Planck
time $T_{pl}$ as the duration of quantum gravitational
interaction. This leads to the result that this interaction is
mediated by the light velocity; a result which is not consistent
with the huge mass $M_{pl}$ of mediating particle. But, it is
easily seen that $T_{pl}$ plays no fundamental role in this game.
This is because, $T_{pl}$ is related to $L_{pl}$ merely by the
classical gravity factor $c$ which plays no role in the quantum
gravity sector. This is not the case for $M_{pl}$ and $L_{pl}$,
because they are related not only by $c$ but also by $\hbar$ which
connects the classical and quantum domains. $M_{pl}$ and $L_{pl}$
are independent factors of quantum gravity, just like momentum and
position in quantum mechanics. $T_{pl}$ in this game is a
redundant factor which plays no important role. }. Since we assume
the quantum rules are just valid in this domain then according to
Heisenberg uncertainty relation the huge mass of $M_{pl}$ implies
that quantum gravitational interactions have ultra-short range
$L_{pl}$, as desired in consistent with the assumption that they
are confined to the planckian domain.

The Schwarzschild radius of the planck mass is obtained as
\begin{equation}
R_s=2L_{pl},
\end{equation}
which means the range of quantum gravitational interaction
$L_{pl}$, is inside the Schwarzschild radius. In fact, the quantum
particle with mass $M_{pl}$ fits into the Schwarzschild radius
because its Compton wavelength $L_{pl}$ is smaller than $R_s$.
Therefore, the quantum gravitational interactions are always
hidden from the domain of classical gravity. The appearance of
Planck size Schwarzschild radius is in complete consistency with
GCP, provided we assume the existence of microscopic wormholes
with Planck size throat to mediate Planck mass particles as the
quantum gravitational interaction. The wormholes with Planck size
throat can hide the Planck range quantum gravitational interaction
from outside the horizon. In fact, the throat of these wormholes
may define the zero point length of $g_{\mu \nu}$, namely the
Planck length, below which $g_{\mu \nu}$ is not defined. This
throat is the boundary of the government of the classical gravity!

In conclusion, it seems a quantum gravitational interaction takes
place as follows: whenever two particles are accelerated toward
huge energies to collide each other at a ultra-short distance of
quantum gravity regime, a virtual microscopic wormhole of Planck
size is created and the particles interact with each other quantum
gravitationally, through this wormhole, by exchanging virtual
Planck mass particles. \footnote{The virtual wormholes or
particles mediating quantum gravitational interactions are created
according to Heisenberg uncertainty principle $c M_{pl} L_{pl}
\sim \hbar$, which governs at the throat domain.}.

The most important point is the emergence of a new definition for
the mass. If gravity has two different features at over-planckian
and planckian domain, then there are two different definition of
mass in these domains: { \it classical gravitational mass} CGM and
{ \it quantum gravitational mass} QGM. Since classical gravity is
a long range interaction, then CGM is defined in large scale. But,
quantum gravity of the type discussed in this paper is a
ultra-short range interaction. Then, QGM have to be defined at
ultra-short distance of planck length. When two particles are at a
distance larger than the Planck length ( $r\gg L_{pl}$ ) in
$g_{\mu \nu}$ background, no microscopic wormhole and hence no
quantum gravitational interaction takes place and the CGM of
particles interact classically through the smooth geometry $g_{\mu
\nu}$. However, when they are very close at Planck length a
virtual wormhole with the same size is created so that there is no
notion of distance  between the particles at the throat. This is
because, the appearance of this wormhole makes the 3-geometry to
be non-simply connected. In this case the QGM of particles
interact quantum gravitationally through the throat and there is
no classical gravitational interaction between CGM of particles
because there is no meaningful notion of distance $r$ at the
throat. This clearly explains why GCP forbids the simultaneous
classical and quantum gravity descriptions.

\section{Hierarchy problem can solve the cosmological constant problem}

Cosmological constant and hierarchy problems are the most
outstanding problems in the high energy physics. There is a rich
literature about different solutions for both problems. Recently
proposed brane world gravity with extra dimensions is understood
to successfully solve the hierarchy problem \cite{Br}. As yet,
however, it could not solve the cosmological constant problem in a
successful way. This rises a very important {\it new} problem:
{\it How a fundamental theory can solve the hierarchy problem
without solving the cosmological constant problem?} Conventional
wisdom states that if a theory ( e.g. brane gravity ) is capable
of solving the hierarchy problem in a very subtle way, it has to
be of the same capability to solve the cosmological constant
problem. Therefore, it seems a fundamental quantum gravity theory
is the one that in which both problems can be solved at once. We
shall present a mechanism, based on the present idea of quantum
gravity, by which both problems are properly addressed, at once.
In explicit expression: we show that the solution for the
cosmological constant problem is nothing but the hierarchy
problem!

The so-called cosmological constant problem arises in the
following way
\begin{equation}
\Lambda_{eff}=\Lambda_0+{\cal O}(M_{pl}^4), \label{3}
\end{equation}
where $\Lambda_{eff}$ is a combination of bare cosmological
constant $\Lambda_0$ and quantum field theory contributions of the
order of $M_{pl}^4$. These contributions arise due to the vacuum
fluctuations of quantum fields. They can be interpreted as
classical effects because they are associated with the properties
of the vacuum. Therefore, they couple to classical gravity $g_{\mu
\nu}$, through Einstein equation (\ref{1}). Due to huge values of
these contributions $\sim M_{pl}^4$ we face with the so-called
cosmological constant problem, namely a large difference between
the observational bound and theoretical predictions on the value
of cosmological constant. On the other hand, the hierarchy problem
is to explain why Planck scale is enormously larger than
electroweak scale.

A wise insight to both problems reveals that they have a same
feature in common, namely the Planck mass. In other words,
$M_{pl}$ joints the cosmological constant and hierarchy problems.
Therefore, the key solution of both problems certainly lies in the
physics of Planck scale. The Planck scale physics in this paper is
nothing but the ultra-short range quantum gravity. This means the
present idea of quantum gravity and the corresponding wormhole
structure is the key to resolve the enigmatic situation concerning
the cosmological constant and hierarchy problems. To show this, we
notice that the quantum gravity features $M_{pl}$ and $L_{pl}$ in
the present idea provides us with an energy density of the order
of $M_{pl}L_{pl}^{-3} \sim M_{pl}^4$ in the throat domain. But,
this is surprisingly the same vacuum energy density associated
with the quantum field contributions to the effective cosmological
constant in (\ref{3}). Therefore, the microscopic wormholes can
act as drainpipes to evacuate the large contributions $M_{pl}^4$
from classical gravity domain to elsewhere\footnote{A similar idea
has already been proposed to solve the cosmological constant
problem \cite{Co}. In this idea our hot universe is assumed to be
in contact with other large and cool universes by Planck size
wormholes to vanish the effective cosmological constant. }. What
is then left is the bare cosmological constant which can in
principle be adjusted to satisfy the present observations. In this
way, the main purpose of the existence of ultra-short range
quantum gravity is to set the effective cosmological constant to
zero by its wormhole structure. In quantum field theory language,
these wormholes play the role of counter terms to exactly cancel
out the contributions $M_{pl}^4$ from the $g_{\mu \nu}$ universe.

However, it is very important to note that the essential condition
for the occurrence of these microscopic wormholes is the existence
of Planck scale energy $M_{pl}$, not electroweak energy scale
$M_{EW}$. This means that the hierarchy between $M_{EW}$ and
$M_{pl}$ is a natural ( perhaps anthropic ) set-up to solve the
cosmological constant problem. The electroweak energy scale is
responsible for symmetry breaking via Higgs field, whereas the
Planck scale energy is responsible for vanishing the effective
cosmological constant via microscopic wormholes.

\end{document}